\documentclass[twocolumn,aps,floats,epsf,prb]{revtex4}

\usepackage{graphicx}

\begin{document}

\newcommand{\be}{\begin{equation}}
\newcommand{\ee}{\end{equation}}
\newcommand{\bn}{\begin{eqnarray}}
\newcommand{\en}{\end{eqnarray}}
\newcommand{\ii}{\'{\i}}

\title{Orbital-selective Mottness in layered iron oxychalcogenides: 
The case of Na$_{2}$Fe$_{2}$OSe$_{2}$}

\author{L. Craco,$^1$ M.S. Laad,$^2$ and S. Leoni$^3$}

\affiliation{$^1$Instituto de F\ii sica, Universidade Federal de 
Mato Grosso, 78060-900, Cuiab\'a, MT, Brazil \\
$^2$Institut Laue-Langevin, 6 Rue Jules Horowitz, 38042 Grenoble Cedex,
France \\
$^3$Physical Chemistry, Technical University Dresden,
01062 Dresden, Germany}

\date{\rm\today}

\begin{abstract}
Using a combination of local-density approximation (LDA) and dynamical 
mean-field theory (DMFT) calculations, we explore the correlated 
electronic structure of a member of the layered iron oxychalcogenide 
Na$_{2}$Fe$_{2}$OSe$_{2}$.  We find that the parent compound is a 
multi-orbital Mott insulator. Surprisingly, and somewhat reminiscent 
of underdoped high-$T_{c}$ cuprates, carrier localization is found to 
persist upon hole doping because the chemical potential lies in a gap 
structure with almost vanishing density of states.  On the other hand, 
in remarkable contrast, electron-doping drives an orbital-selective 
metallic phase (OSMP) with coexisting pseudogaped (Mott-localised) 
and itinerant carriers. These remarkably contrasting behaviors in 
a single system thus stem from drastic electronic reconstruction caused 
by large-scale transfer of dynamical spectral weight involving states 
with distinct orbital character at low energies, putting the 
oxychalcogenides neatly into the increasingly visible tendency of 
Fe-based systems as ones in OSM phases. We detail the implications 
that follow from our analysis, and discuss the nature and symmetries 
of the superconductive states that may arise upon proper doping or 
pressurising Na$_{2}$Fe$_{2}$OSe$_{2}$.    
\end{abstract}


\maketitle

\section{Introduction}

Recent finding of unconventional high-T$_{c}$ superconductivity (HTSC)
in Fe arsenides and selenides have reinvigorated the HTSC revolution.
In particular, the outstanding and vigorous debates concerning the 
normal state giving way to HTSC in cuprates have also produced their 
mark upon the discussion of mechanism(s) of HTSC in Fe-based 
superconductors (FeSC): Does HTSC result as a pairing instability 
of a conventional Landau-Fermi liquid (LFL),~\cite{chub} or as one 
of a non-LFL, akin to the cuprates.~\cite{pwa} Clearly, one possible 
way to resolve this very important issue is to investigate as many 
members of the Fe-based family as possible.

To date, both, normal paramagnetic and antiferromagnetically ordered 
phases in various members have been extensively studied. It has slowly
emerged that, in their normal, paramagnetic state the FeSC (both pnictides 
and chalcogenides) increasingly fall into the bad metal category.  Usually, 
in correlated systems, the generic picture is known by now to be one where 
a high-$T$ bad-metal undergoes a $T$-dependent crossover to a correlated 
or heavy LFL at low temperature $(T)$. This is due to the development of 
a severely renormalised lattice coherence scale, driven by increasing 
relevance of electronic correlations as $T$ reduces. We will refer to 
this behavior as ``standard'' for correlated metals, and thus any 
fundamental deviation from this generic behavior will be termed 
``anomalous''.  According to this definition, increasing number of 
Fe-based compounds fall into the anomalous class, in the sense that 
LFL behavior is seemingly not recovered at low $T$ when superconductivity 
(SC) is suppressed by appropriate perturbations. In 
SmFeAsO$_{1-x}$F$_{x}$,~\cite{boebinger1} for example, destruction of SC 
by high magnetic fields reveals a low-$T$ insulator-like state as in 
high-$T_{c}$ cuprates.~\cite{hill} NdFeAsO$_{1-x}$F$_{x}$ shows similar 
behavior upon irradiation-induced disorder.~\cite{NdFe} In both cases, 
the residual resistivity, $\rho_{r}(T)\simeq -$log$T$, suggesting an 
anomalous field- or radation (disorder) driven bad-metal-insulator 
transition atlow $T$ in the normal state, and that the {\it normal} 
state might thus not be as normal as thought. SC even arises directly 
from an even more incoherent semiconductor-like normal state in 
FeSe$_{1-x}$Te$_{x}$.~\cite{FeSeTe} Optical and spectroscopic studies 
in 122-iron selenide~\cite{opt122Se} superconductors show large-scale 
spectral weight transfer (SWT) as a function of temperature across the 
magnetic and superconducting instabilities, a fingerprint of 
Mottness.~\cite{Mness} Both $\rho(T)\simeq T$ and bad metallicity 
above $T_{c}$ or the N\'eel temperature ($T_N$) are features shared 
along with other non-LFL metals close to a Mott 
instability.~\cite{sachdev} Thus, the non-LFL metallic state in the 
above cases strongly points toward a fundamental role for Mott 
localization.~\cite{our122,fesehexag,sioxy} 

\begin{figure*}[t!]
\centering\includegraphics[width=5.6in,angle=0]{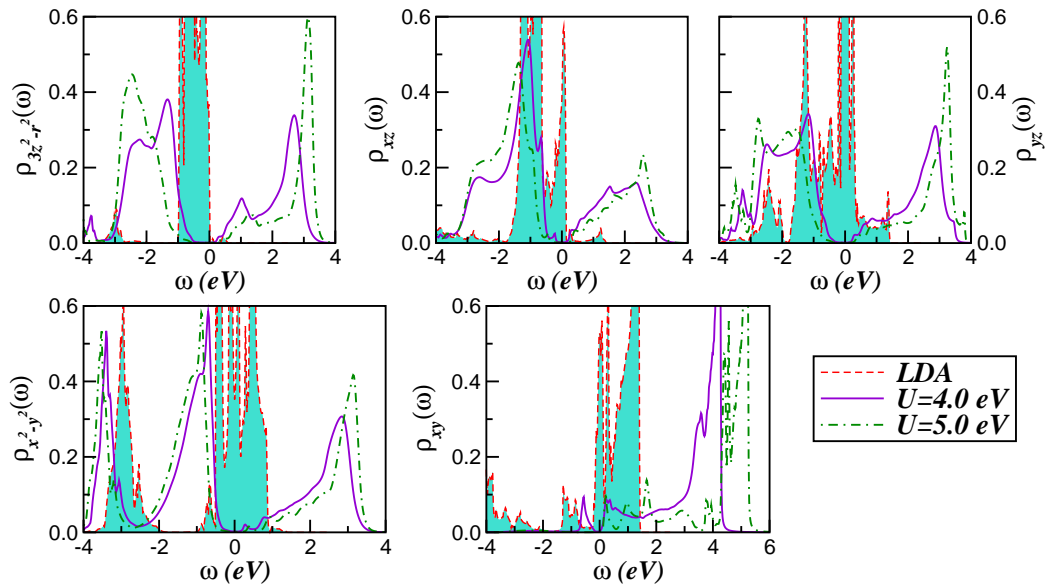}
\caption{(Color online)
Orbital-resolved LDA and LDA+DMFT (two values of $U$ and fixed 
$J_{H}=0.7$~eV) density-of-states (DOS) for the Fe $d$ orbitals of 
Na$_{2}$Fe$_{2}$OSe$_{2}$. Notice the narrow bands in the LDA DOS. 
Compared to the LDA results, large spectral weight transfer along 
with Mott localization is visible in the LDA+DMFT spectral functions.}
\label{fig1}
\end{figure*}

In this context, recent finding of insulating, AF ground state in layered 
Fe-oxychalcogenides~\cite{sioxy,kab,cavaoxy,he,fei} is interesting, 
since it confirms theoretical suggestions that Mott-insulating parent 
compounds in the FeSC systems could be found by increasing the ratio 
of the interaction-to-hopping $(U/W)$ beyond a critical value for a 
Mott transition.~\cite{basksi,fesehexag}  Evidence for Fe-layered 
compounds with different intergrowth structures is very interesting: 
Its generic effect is to reduce the one-electron bandwidths~\cite{sioxy} 
($W$) by employing two-dimensional secondary building units, making it 
possible to realize almost isolated two-dimensional Fe-layers. The 
newly discovered Fe-oxychalcogenides thus help to sharpen the 
fundamental debate~\cite{si-nature} on the degree of electronic 
correlations in FeSC in general. Finding of large local moment value 
on Fe, ${\bf M}_{Fe}\simeq 2.23\mu_{B}$,~\cite{cavaoxy} along with 
large activation energies~\cite{kab,cavaoxy,sioxy,he} suggests 
strong electronic localization. The consequences of Mottness upon
physical responses of oxychalcogenides has, however, not received the 
attention it deserves.

Here we adopt a Mottness view with incorporation of sizable multi-orbital 
(MO) correlations in the Fe $d$ shell. As shown below, all five $d$-bands 
must be kept at a minimally realistic level in order to satisfactorily 
resolve the Mott insulating and bad metallic regimes obtained, 
respectively, in pure and doped compounds. In Mott insulators, sizable 
electronic correlations drives new physical effects upon (electron, hole) 
doping. They can induce a pseudogap regime referred to above,
where the chemical potential lies in an energy region of vanishing
density of states (DOS),~\cite{phillips} or orbital-selective (OS)
incoherent states, naturally yielding co-existent insulating and bad 
metallic states as in cuprates or in increasing number of 
pnictides.~\cite{boebinger,simonson,tokura} In this work, we use the
local-density approximation plus dynamical mean-field theory
(LDA+DMFT)~\cite{kot-rev} to study these issues.  We also use these 
results to discuss the influence of Mottness on issues mentioned above 
in lightly doped Na$_{2}$Fe$_{2}$OSe$_{2}$ and follow this up with 
specific predictions which can be tested in future experiments.

\section{Results and discussion}

The crystal structure of Na$_{2}$Fe$_{2}$OSe$_{2}$ (space group 
$I4/mmm$) is built from alternate stacking of Fe$_{2}$OSe$_{2}$ blocks 
and double layers of Na along c-axis. In the Fe$_{2}$OSe$_{2}$ unit, 
Fe$^{2+}$ ($d^6$ configuration) ion is located between oxygen atoms, 
forming a square-planar layer, which is an anti-configuration to with respect
to the CuO$_2$ layer of high-$T_c$ cuprates. Using the cell parameter 
values,~\cite{he} we performed LDA calculations for Na$_{2}$Fe$_{2}$OSe$_{2}$ 
within the linear muffin-tin orbitals (LMTO) scheme,~\cite{ok} 
in the atomic sphere approximation. Self-consistency was reached by 
performing calculations with 163 irreducible {\bf k}-points. The radii 
of the atomic spheres were chosen as $r$=2.512~(Fe), $r$=3.05~(Na), 
$r$=3.422~(Se) and $r$=1.99~(O) a.u. in order to minimize their 
overlap. Within LDA, the one-electron part of the Hamiltonian reads 
$H_{0}=\sum_{{\bf k},a,\sigma}
\epsilon_{a}({\bf k})c_{{\bf k},a,\sigma}^{\dag}c_{{\bf k},a,\sigma}$,
where $a=(3z^{2}-r^{2},xz, yz,x^{2}-y^{2},yz)$ label the (diagonalized 
in orbital basis) five Fe $3d$ bands, which are the only ones we retain, 
since the non-$d$-orbital DOS have negligible or no weight at $E_{F}$.

\subsection{Electronic structure}  
 
In Fig.~\ref{fig1} we show the LDA one-electron band structure, 
confirming that the active electronic states involve the Fe $d$ carriers.
A sizable reduction (O$(30)\%$) of the average LDA bandwidth $(W_{LDA})$ 
relative to that of tetragonal FeSe,~\cite{fesehexag} induced by 
hybridisation with oxygen atoms and distorting effects of the extra 
Na layers, is obtained. As shown previously, FeSe is already a bad 
metal close to a Mott insulator,~\cite{fesehexag} 
and the significantly smaller $W_{LDA}$ for Na$_{2}$Fe$_{2}$OSe$_{2}$ then 
naturally implies favoring a Mott insulator in oxychalcogenides.  The 
MO-driven anisotropies in the LDA band structure are also manifested 
in Fig.~\ref{fig1}. As is common to all FeSC, the $3z^{2}-r^{2},xy$ 
orbitals exhibit the bonding/antibonding splitting characteristic of 
the tetragonal unit cell. These bands, almost totally band-gapped near 
the Fermi energy ($E_{F}$) in the FeSC systems go over into highly 
orbitally polarized, narrow bands in Na$_{2}$Fe$_{2}$OSe$_{2}$.  A similar 
feature is found for the $xz$ orbital.  Moreover, the real crystal-field 
splitting also severely renormalizes the LDA orbital occupancies 
(promoting enhanced orbital polarization) of the $3z^{2}-r^{2},xz,xy$ 
sectors as well. In another important distinction to the FeSC, the 
(strong) crystal-field splitting also lifts the degeneracy of the 
$xz,yz$ orbitals, leaving an AF ground state at low-$T$ without any
tetragonal-to-orthorhombic structural phase transition. It thus 
follows that the novelties found in FeSCs, relating to electronic 
nematic instabilities in the tetragonal phase near the borderline of 
structural and magnetic transitions, will not play an active role in 
oxychalogenides. Thus, oxychalcogenides should show very different 
responses upon carrier doping: elucidating precisely this aspect 
with a well-controlled approach will be our aim in this work.

We now discuss our LDA+DMFT results obtained within the $d^6$ configuration 
of the Fe$^{2+}$ in Na$_{2}$Fe$_{2}$OSe$_{2}$. Starting with the five Fe 
$d$-orbitals in LDA, we use MO-DMFT~\cite{our122} to derive the correlated 
$d$-band spectral functions. We use the MO iterated perturbation theory as 
an impurity solver for DMFT.~\cite{ePAM} This perturbative, many-particle 
scheme has a proven record of recovering correct LFL behavior, and 
bad-metallicity in correlated oxides. In the MO problem: 
$H=H_{0} + H_{int}$ with 

\bn
\nonumber
H_{int} &= & U\sum_{i,a} n_{i,a\uparrow}n_{i,a\downarrow} \\ \nonumber
&+& \sum_{i,a \ne b} [U' n_{i,a}n_{i,b} - J_{H} {\bf S}_{i,a} \cdot {\bf S}_{i,b}] \;,
\en
$U'\equiv U-2J_H$ [$U~(U')$ is the intra- (inter-) 
orbital Coulomb repulsion] and $J_H$ is the Hund's rule 
coupling.~\cite{kot-rev} We choose values of $U=4.0$~eV, $J_{H}=0.7$~eV 
as employed in our earlier works.~\cite{our122,ouroptc} Though not 
determined {\it ab initio} our parameter choice is consistent with our 
earlier and other theoretical works and with our underlying view that 
Fe-pnictides and -chalcogenides are better regarded as sizably correlated 
metals.~\cite{ouroptc,lili} 

\begin{figure*}[t!]
\centering\includegraphics[width=5.5in,angle=0]{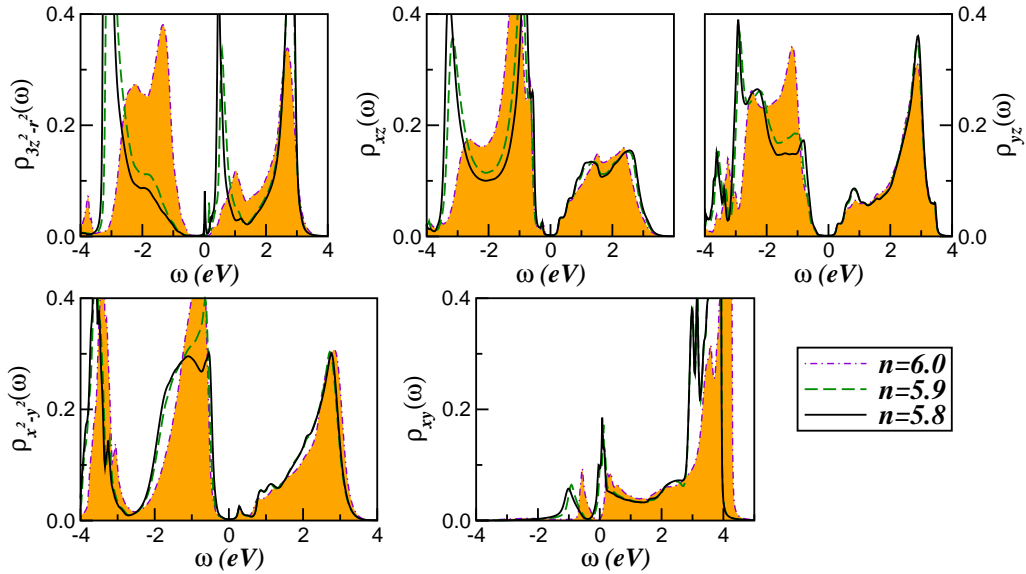}
\caption{(Color online)
Hole doped LDA+DMFT ($U=$4.0~eV, $J_H=$0.7~eV) spectral 
functions for the Fe $d$ orbitals of Na$_{2}$Fe$_{2}$OSe$_{2}$. 
As in cuprates,~\cite{phillips} strong hole localization is 
found in the lightly doped regime: only narrow gap states 
are formed in the slightly more delocalized $3z^2-r^2,xy$ orbitals.}
\label{fig2}
\end{figure*}

\subsection{Mott-insulating phase}

In the case of undoped Na$_{2}$Fe$_{2}$OSe$_{2}$, our LDA+DMFT results,
shown in Fig.~\ref{fig1} exhibit a clear Mott insulating gap in the 
one-particle spectral function.  Several interesting features are 
clearly stand out: (i) The Mott gap is orbital-dependent, i.e, 
intrinsically anisotropic. (ii) Examination of the orbital-resolved 
spectral functions reveal a behavior hitherto known to FeSC 
systems. Namely, all orbitals are partially populated due to $U'$-induced 
dynamical inter-orbital entanglement,~\cite{ePAM} while the total 
orbital population is integral, as it should be for undoped $(\delta=0)$ 
Na$_{2}$Fe$_{2}$OSe$_{2}$. As seen in Fig.~\ref{fig1}, strong dynamical 
MO correlations originating from $U,U'$ and $J_{H}$ lead to sizable 
spectral weight redistribution over large energy scales and the formation 
of a severely reconstructed (compared to LDA) correlated electronic 
structure.  This feature is characteristic of multiband Mott systems, 
with concomitantly modified upper and lower Hubbard bands at high-energies: 
These latter features are related to the strongly coupled spin-orbital 
local moments defining a Mott insulator without long-range orbital or 
magnetic order.  Though our results pertain to $T=0$, they are thus 
formally a valid description of the high-$T$ Mott insulator state
described before.
 
\subsection{Filling-controlled electronic transition}

What happens upon carrier doping?  Even though no data exists, 
the generic appearance of novel metallic states with low energy 
pseudogaps, and the instabilities of such states to unconventional order,
and, in particular, to HTSC, in a wide variety of other correlated matter
makes this a very important question to inquire about.  As alluded to 
in the introduction, our aim here is to build upon the strengths of 
correlated electronic structure modelling to {\it predict} the effect of
carrier doping.  In particular, based on explicit calculations, we will 
present a set of predictions which could be tested in future experimental 
work.

\begin{figure*}[t!]
\centering\includegraphics[width=5.5in,angle=0]{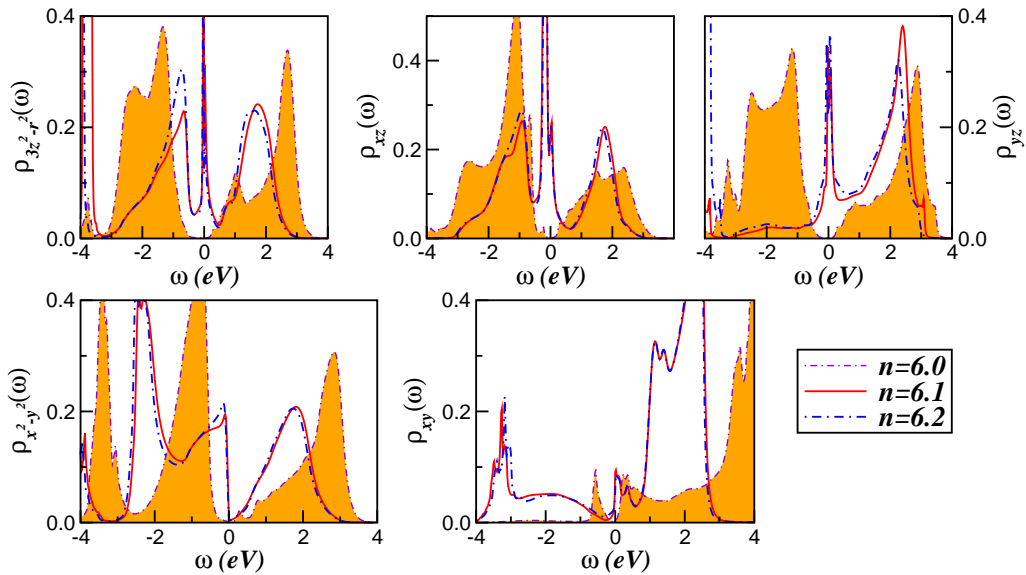}
\caption{(Color online)
Effect of electron doping on the LDA+DMFT ($U=$4.0~eV, $J_H=$0.7~eV) 
spectral functions for the Fe $d$ orbitals of Na$_{2}$Fe$_{2}$OSe$_{2}$. 
Electron delocalization and appearance of narrow quasiparticle 
resonances near the Fermi energy is seen within the $3z^2-r^2,xz,yz$ 
channels. Selective-localization and pseudogap features are the 
fingerprints of electronic correlations within the $x^2-y^2,xy$ orbitals.}
\label{fig3}
\end{figure*}

In Fig.~\ref{fig2} we show the changes in the correlated electronic 
structure upon hole doping ($n\equiv 6+\delta$, with $\delta<0$) the
Mott insulator. An intriguing observation is that the Mott
localisation-delocalisation phase transition does {\it not} occur 
at small doping. However, as $|\delta|$ increases, a selective-Mott 
state develops, in which the $xz,yz,x^2-y^2$ spectral functions show 
behavior of a Mott insulator with vanishing DOS at $E_F$, while the 
$3z^2-r^2,xy$ orbitals now show (selective) metallic behavior, 
characterized by the presence of incoherent in-gap states at $E_{F}$. 
This is a clear demonstration of strong Mottness~\cite{phillips} 
and an orbital-selective Mott transition (OSMT) in 
Na$_{2}$Fe$_{2}$OSe$_{2}$.  What is the origin of these features?
In a MO system like the oxychalcogenides, strong (incoherent) 
scattering between different carriers in orbital states split relative to 
each other due to the specific crystal field (see discussion of this aspect 
above) leads to two main effects: (i) It leads, via static-Hartree 
contributions (from the static part of the orbital-dependent self-energies)
to an-orbital-dependent shifts of the $d$-bands relative to each other, and
(ii) strong dynamical correlations due to sizable $U,U',J_{H}$ cause 
appreciable SWT over large energy scales, from high to low-energy, upon 
carrier doping.  This second feature leads to a drastic modification of 
the spectral lineshapes.  In full accord with our earlier qualitative 
discussion, the increase of the effective $U/W$ ratio in the present 
case relative to FeSe and Fe-arsenides leads to increased low-energy 
incoherence, as the pronounced pseudogap {\it and} very broad low-energy 
spectral features clearly shows.  Microscopically, strong incoherent 
scattering, arising from co-existence of Mott-localized and bad metallic 
states, leads to an almost complete suppression of the LFL quasiparticles 
and the emergence of an incoherent (pseudogaped) spectra, reminiscent of 
what is seen in cuprate oxides. Microscopically, infrared 
LFL behavior (narrow Kondo resonance in DMFT) in the $3z^2-r^2,xy$ orbitals 
is almost extinguished by strong scattering between the Mott-localized
$xz,yz,x^2-y^2$ and the quasi-itinerant $3z^2-r^2,xy$ components of the 
(DMFT) matrix-spectral function, due to sizable $U',J_{H}$, and is 
a clear manifestation of the OSMT in the five-band Hubbard model we use. 

Since there is no particle-hole symmetry in the system, it is interesting to 
inquire as to the effects of electron doping ($n\equiv 6+\delta$, $\delta>0$) 
in Na$_{2}$Fe$_{2}$OSe$_{2}$. In particular, we want to study if electron
doping is qualitatively different; i.e, whether incoherent non-LFL behavior
still survives in the infrared, or if coherent LFL response is favored. 
Fig.~\ref{fig3} exhibits the answer: The $x^{2}-y^{2},xy$ spectral functions
show clear pseudogap behavior, while, very interestingly and in stark contrast 
to the hole-doped case, the $3z^{2}-r^{2},xz,yz$ spectral functions now show 
narrow LFL-like quasiparticles at $E_{F}$. Hence, electron doped 
Na$_{2}$Fe$_{2}$OSe$_{2}$ is predicted to 
lead to a more coherent (in the LFL sense in the infrared) state than 
its hole-doped counterpart above. The situation seems to be somewhat 
similar to the 122-FeAs systems,~\cite{dressel} where both coherent 
LFL-like and strongly incoherent non-LFL responses, dependent on the 
{\it type} and extent of doping, are seemingly observed.  We point out, 
however, that such comparisons are inevitably fraught with danger, 
especially since the $xz,yz$ orbital degeneracy and the soft orbital 
fluctuations characteristic of the 122-FeAs systems do {\it not} play 
a role in the Fe-oxychalcogenides. 

\begin{figure}[t]
\includegraphics[width=\columnwidth]{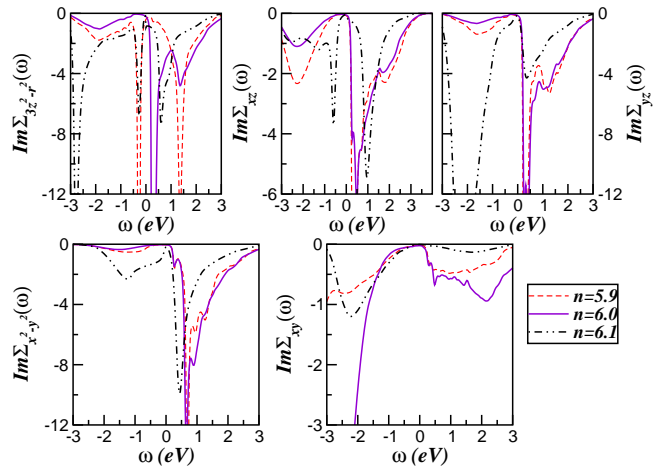}
\caption{(Color online)
Orbital-resolved imaginary parts of the self-energies for the 
Fe $d$ orbitals of stoichiometric and (electron/hole) doped 
Na$_{2}$Fe$_{2}$OSe$_{2}$ with $U=4.0$~eV, $J_{H}=0.7$~eV. 
Notice the evolution of the sharp pole in the self-energies 
near $E_F$ across the doping-induced selective-Mott delocalization.}
\label{fig4}
\end{figure}

The electron-hole asymmetry and selective-Mott transition in 
Na$_{2}$Fe$_{2}$OSe$_{2}$ is further visible in orbital resolved 
self-energies, $\Sigma_{a\sigma} (\omega)$, shown in Fig.~\ref{fig4}.  
Noticeable features include: (i) The insulating state is manifested 
by a sharp pole in Im$\Sigma_{a\sigma} (\omega)$ as well by 
divergent Re$\Sigma_{a\sigma} (\omega)$ (not shown) close to $E_F$. 
(ii) In the doped regime, only the higher (in energy) $xy$ orbital  
shows LFL (Im$\Sigma_{a\sigma} \approx -\omega^{2}$) form at small 
$\omega$. On the other hand, (iii) the imaginary part of the 
self-energy of the  $3z^{2}-r^{2}$ orbital at $E_F$ is finite (i.e., 
Im$\Sigma_{3z^{2}-r^{2}\sigma} (\omega=0)<0)$.  As is known, such 
behavior is caused by strong scattering between Mott localized and 
quasi-itinerant electronic states in the MO-DMFT problem, which 
maps onto an {\it effective} spinful Falicov-Kimball model in the 
local limit. These results imply that underdoped oxychalcogenides 
should be located in pseudogap regime and the metallic state obtained 
by the filling-controlled Mott transition is unconventional.~\cite{our122}  
The origin of this unconventional metal is the lattice orthogonality 
catastrophe that occurs due to orbital-selective blocking of the 
coherent motion of the doped hole in the DMFT due to sizable $U',J_{H}$ 
in the MO Hubbard model.

\begin{figure}[t]
\includegraphics[width=\columnwidth]{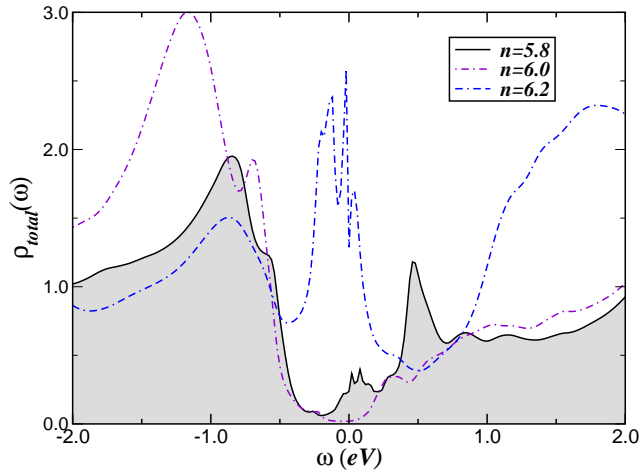}
\caption{(Color online)
Total density-of-states for the Fe $d$ orbitals of stoichiometric 
and (electron/hole) doped Na$_{2}$Fe$_{2}$OSe$_{2}$ with 
$U=4.0$~eV, $J_{H}=0.7$~eV. Notice the modification of the LDA+DMFT 
spectra from a Mott insulator $(n=6)$ to an almost totally incoherent 
electronic state ($n=5.8$) due to large-scale transfer spectral weight.
Selective microscopic coexistence of coherent-incoherent low-energy 
electronic states is predicted for electron-doped $(n=6.2)$ 
Na$_{2}$Fe$_{2}$OSe$_{2}$.}
\label{fig5}
\end{figure}

Armed with the above results,~\cite{ouroptc} we now attempt to predict 
the detailed physical response in the paramagnetic state of 
Na$_{2}$Fe$_{2}$OSe$_{2}$.  In Fig.~\ref{fig5}, we show the total 
LDA+DMFT spectral functions. Clear Mott insulator features are visible 
for $\delta=0$, and we propose that future photoemission and X-ray 
absorption spectroscopy results, which probe one-electron subtraction 
and addition spectra, can be directly compared with these: In particular, 
a broad incoherent peak below $-0.8$~eV should be seen in both pure and 
doped cases. Additionally, drastic modification of the LDA+DMFT spectra 
with $\delta\neq 0$ is clearly visible. Our results for the hole- and 
electron-doped spectral functions ($n=5.8,6.2$) reveal an clear 
differences, as found above: For hole-doped case, we find an almost 
totally incoherent spectral response due to the almost total blocking 
of coherent one-electron dynamics due to the lattice ``orthogonality 
catastrophe''as noted above. For electron doping, however, a small 
quasicoherent LFL component is visible, notwithstanding large-scale 
dynamical spectral weight transfer common in both cases. These are 
stringent tests for our proposal, and experimental verification should 
place it on solid ground.  More distinguishing tests would be 
characteristic signatures in transport: For example, incoherent 
bad-metallicity with clear non-LFL $T$-dependence of the resistivity, 
$\rho(T)\simeq T^{n}$ with $0\leq n<1.5$ should be seen in the 
hole-doped non-LFL cases, while much more conventional LFL-like 
resistivity, $\rho(T)\simeq T^{2}$ should be seen in the 
electron-doped cases. 

One of our central results is the finding of co-existing orbital-selective
incoherent and heavily dressed but coherent charge carriers. This behavior 
is reminiscent of what is found in the pseudogap regime in the underdoped 
HTSC cuprates.  However, in cuprates, this differentiation of electronic 
states occurs in momentum space: Carriers in the nodal region are more
quasicoherent than ones along the antinodal direction, which remain 
incoherent and in Mott localised states.  It has been demonstrated, 
using a cluster-to-orbital mapping~\cite{liebsch} that this 
nodal/antinodal differentiation in one-band Hubbard models is the 
momentum-space analog~\cite{ferrero} of the orbital-selective Mott transition 
in models with several active orbitals. This is also borne out by our
findings here, where the frequency dependence of the imaginary part of 
a subset (of orbitals self-energies) in Fig.~\ref{fig4} show Mott 
insulator features, while others show quasicoherent behavior. Viewed 
in terms of the above mapping, an analogy can be drawn with the 
observations of cellular-DMFT studies,~\cite{ferrero,kot-rev}
wherein similar features within the two-site cluster-DMFT were found: 
Suppressed coherence of antinodal quasiparticles originated from Mott 
physics while the nodal states remained {\it protected} against Mott 
localisation. Selective-localization in momentum (high-$T_c$ cuprates) 
or in orbital (FeSC) space thus seems to be a common element that 
provides a common, microscopic origin for the suppression of LFL 
coherence and emergence of non-LFL normal state properties of 
unconventional metals.~\cite{hill} 

\subsection{Mechanisms of unconventional superconductivity}

Finally, motivated by a host of studies~\cite{mazin,laad,ronny,chub} which 
propose the possibility and gap-function symmetry of superconducting 
states based on the nature (LFL or non-LFL) of the normal state and the 
nature of the renormalised Fermi surface(s), we attempt to carry out a 
qualitative program in our case. Specifically, we will focus on two 
normal state features: (i) The LFL or non-LFL character of the normal 
state(s) found above, and (ii) the natures of the renormalised Fermi 
surfaces to explore the range of superconduting state(s) that may result.  

$(i)$ {\it Hole doping}: Since LFL picture has been totally extinguished 
in the normal state for hole doping, it follows that the instability to 
any unconventional superconducting state {\it cannot}, by definition, 
proceed via the conventional route of BCS-like pairing of (heavily 
dressed or otherwise) long-lived quasiparticles: It is then much more
likely, as occurs in critical metals,~\cite{laad} that SC pairing 
occurs as a result of the relevance of the inter-site, inter-orbital 
two-particle residual interactions in the incoherent metal. As 
worked out for the multiband situation characteristic of Fe-based 
systems, such a residual interaction should involve pair-hopping 
electronic processes with MO character. In the hole-doped case above, 
with metallic $xy,3z^{2}-r^{2}$ states crossing $E_{F}$, we expect 
pairing to primarily involve solely the $xy$ band or both 
$xy,3z^{2}-r^{2}$ states if the Fermi surface has appreciable warping 
along $k_{z}$. Once SC has stabilised
in the $xy$ band, an interband proximity effect characteristic of multiband 
SC~\cite{rice} must eventually take over to give full three-dimensional SC. 
Since the two-particle hopping is expected to acquire the frustrated 
form factor $\gamma(k)=($cos$k_{x}+$cos$k_{y})+\alpha$cos$k_{x}$cos$k_{y}$
(since the residual interaction scales as the second power of the one-electron
hopping),~\cite{laad} the SC gap function may or may not have in-plane nodes,
depending on whether the {\it renormalised} $xy$ band dispersion intersects
the superconducting gap function. 

$(ii)$ {\it Electron doping}: Since more quasicoherent normal state behavior
sets in for electron-doping, a more conventional analysis for superconducting 
pairing is mandated.  Remarkably, since LFL-like behavior in the infrared 
now obtains for the $xz,yz,3z^{2}-r^{2}$ bands, the situation that emerges 
is similar to the one~\cite{mazin,chub,slutko} found for the 
1111-FeAs systems in early itinerant-based studies, though the present 
result places these bands much closer to the Mottness regime, as an 
examination of our results above shows.  Thus, it is possible that 
small additional perturbations might tilt it into the incoherent regime.
This analogy can now be used to directly predict that the superconduting 
state will most probably have in-plane $s_{\pm}$ pair symmetry. Further, 
depending on the extent of the $c$-axis corrugation of the three-dimensional 
(3D) Fermi surface, it will (or will not) have accidental $c$-axis 
nodes.~\cite{laad,mazinsc}

It is thus quite remarkable that {\it both}, the nature and extent of 
orbital selectivity in hole- and electron-doped cases play a critical role 
in choosing the detailed set of conditions which determine the nature and
symmetry of the superconducting state in each cases.  The diversity of 
pairing states in Fe-based systems, by now appreciated to be a 
consequence of the MO nature of these systems, may thus be exposed in Fe 
oxychalcogenides if these can be driven superconducting by appropriate 
pressure, chemical and/or electrical doping.  Whether this fascinating 
state of affairs unfolds in the future only time will tell.
 
\section{Conclusion}

To summarize, we have used LDA+DMFT for a minimally realistic five-band
Hubbard model to perform a detailed study of a doped Mott insulator in
the recently discovered Fe oxychalcogenide.  In particular, considering
Na$_{2}$Fe$_{2}$OSe$_{2}$ as a suitable template, we have carefully analysed
its paramagnetic Mott state, baring the orbital-selective Mott gap as an
effect of multi-orbital Hubbard correlations. Much more interesting behavior 
is predicted when we consider electron- and hole doping of this state: 
Upon the latter, hole localization or incoherence persists because the 
chemical potential lies in a gap region with vanishing density of states, 
or in the low-energy pseudogap, which we ascribe to the blocking of 
coherent motion of doped holes due to a lattice orthogonality catastrophe 
induced by orbital-selective Mottness. On the other hand, in a remarkable 
difference, electron doping this Mott insulator leads to an orbital-selective 
low-energy quasi-coherent states co-existing with pseudogapped states, 
implying low-energy strongly correlated Fermi-liquid state with a small 
quasiparticle weight. These can be directly tested by a combination of 
future spectral and transport measurements.  Finally, based upon the 
reconstructed Fermi surfaces implied by our results as well as upon 
the non-LFL or LFL nature of the metallic phases found here, we 
discuss the nature and symmetry of the possible unconventional 
superconducting states that one may find upon proper doping or 
pressurising the system.  Such studies are called for, and should 
confirm or refute our predictions.

L.C. thanks the Brazilian funding agency CAPES (Proc. No. 002/2012) 
for financial support. Acknowledgment is also made to FAPEMAT/CNPq 
(Projet: 685524/2010).

\end{document}